\documentclass{appolb}
\usepackage{amsfonts}

\usepackage{epsfig}
\usepackage{rotating}
\usepackage{hyperref}
\usepackage{comment,color}
\usepackage{braket}
\usepackage{amssymb,amsmath,bm}
\usepackage{setspace,lipsum}
\usepackage[T1]{fontenc}

\def\be{\begin{equation}}
\def\ee{\end{equation}}
\def\bdm{\begin{displaymath}}
\def\edm{\end{displaymath}}

\input{tcilatex}

\begin{document}

\title{Formation and Deformation of the $\psi(3770)$ \thanks{%
Talk presented by S.~Coito at the ``Excited QCD'' Workshop, Sintra,
Portugal, 7-13 May 2017.} }
\author{S.~Coito$^a$, F.~Giacosa$^{a,b}$ \address{$^a$Institute of Physics,
Jan Kochanowski University, 25-406 Kielce, Poland} 
\address{$^b$Institut f\"ur Theoretische Physik, Johann Wolfgang Goethe-
Universit\"at, 60438 Frankfurt am Main, Germany} }
\maketitle

\begin{abstract}
The form of resonance line-shapes unveils information about its
nonperturbative properties and formation mechanisms. Here, we study the
non-Breit-Wigner energy distribution of the resonance $\psi (3770)$ using an
unitarized effective Lagrangian approach, that includes the effect of the
nearby threshold $D^{+}D^{-}$. Two poles are found in the second Riemann
sheet near the resonance amplitude. We discuss the setting of the free
parameters and possible effects contributing to the signal.
\end{abstract}

\section{Introduction}

The resonance $\psi (3770)$ is listed in Particle Data Group with average
parameters $M=3778.1\pm 1.2$ MeV and $\Gamma =27.5\pm 0.9$ MeV \cite{PDG}.
The state is predominantly a $n^{2S+1}L_{J}=1^{3}D_{1}$ vector charmonium;
it is just above the $D\bar{D}$ hadronic decay channel. In Ref.\ \cite%
{prl101p102004} BES data have shown a clearly non-Breit-Wigner line-shape.
In other data, such as in BaBar \cite{prd76p111105} and KEDR \cite%
{plb711p292}, such distortion is also visible, namely a higher slope on the
right side of the resonance, while on the left energy side the slope appears
to display, in addition, a structure.

One aims to understand the reason for such asymmetries in the line-shape
during the formation of the $\psi(3770)$. Interferences due to the $D\bar{D}$
kinematic background are the most obvious to consider, as shown in Ref.\ 
\cite{prd80p074001}, justifying the higher slope on the right. The
contribution of the $\psi(2S)$ is also taken into account in Refs.~\cite%
{prd86p114013}. Indeed, it is likely that though dominantly a $D$-wave, the $%
\psi(3770)$ is a mixed $^ 3D_1-\ ^3S_1$ state. The inclusion of such effect
in itself is not enough to reproduce the structure of the left side of the
resonance that has been seen in \cite{prl101p102004}, though, within errors,
it is in agreement with the data. In Ref.~\cite{prd88p014010} an estimation
of the non-$D\bar{D}$ hadronic background has been performed, though it
should be residual. Predictions involving $p\bar{p}$ production, leading to
higher cross sections, have been made in Refs.~\cite{prd91p114022,epjc76p192}%
. More within our goal, in Ref.~\cite{plb718p1369} nonperturbative dynamical
effects in the formation of the $\psi(3770)$ are studied in an effective
Lagrangian model including the $\psi(2S)$, $D\bar{D}$ loops and $D-\bar{D}$
rescattering.

In this study, we analyze the dynamical contribution of the $D^{+}D^{-}$
loop to the deformation in the line-shape of the $\psi (3770)$, by employing
an unitarized effective Lagrangian model. Moreover, driven by the suggestion
in Ref.~\cite{prl101p102004}, of a two resonance structure, we study the
poles on the second Riemann sheet. Indeed, similar models have been employed
to light-meson systems where it has been shown that, besides the regular
\textquotedblleft seed\textquotedblright\ pole, extra dynamical poles have been found, alias
the $a_{0}(980)$ \cite{prd93p014002} and the $\kappa (800)$ \cite{npb909p418}%
, leading to deformed line-shapes in the amplitude (for previous work on the
subject, see Ref. \cite{tornqvist}). Similar phenomena is not forbid to
exist for heavy systems.

\section{An Effective Description of $\protect\psi (3770)$}

\subsection{The Lagrangian}

We consider the decay $\psi (3770)\rightarrow D^{+}D^{-}$ of a charmonium
vector to two pseudoscalars. The interacting Lagrangian density $\mathcal{L}%
_{I}$ is defined by 
\begin{equation}
\mathcal{L}_{I}=ig_{\psi DD}\psi _{\mu }\Big(\partial ^{\mu
}D^{+}D^{-}-\partial ^{\mu }D^{-}D^{+}\Big),  \label{lagi}
\end{equation}%
where the fields $\psi $, $D^{+}$ and $D^{-}$ are interacting in the space
with a coupling $g_{\psi DD}$. (An analogous, here omitted, interaction term
couples $\psi $ to $D^{0}\bar{D}^{0}$). This Lagrangian leads to the
amplitude $|\mathcal{M}|^{2}$ for the process $\psi \rightarrow D^{+}D^{-}$: 
\begin{equation}
|\mathcal{M}|^{2}=\frac{4}{3}g^2_{\psi DD}\ p^{2}(s)f(p)\text{ ,}  \label{amp}
\end{equation}%
where $p(s)$ is the relativistic center-of-mass (CM) momentum of $D^{+}D^{-}$%
, with $s$ the CM energy squared, and $f(p)$ is an extra cutoff function
that ensures the convergence of the self-energy (defined in Sec.~\ref{sf})
with the momentum. We use the damping form 
\begin{equation}
f(p)=e^{-2p^{2}/\Lambda ^{2}},  \label{cutoff}
\end{equation}%
where $\Lambda $ is the cutoff parameter. Hence, the model contains two free
parameters, $g_{\psi DD}$ and $\Lambda $. If we assume $\Lambda $ to be
proportional to the inverse of the size of the wave-function, the Fourier
transform of Eq.~\eqref{cutoff} leads to a Gaussian in coordinate space that
models a wave-packet. Therefore, we can estimate the size of our system
using a Schr\"{o}dinger model. Formally, the cutoff function $f(p)$ can be
included in the Lagrangian by rendering it nonlocal \cite{nonlocal};
moreover, even if we use a 3D cutoff, the covariance is satisfied, see
details in\ Ref.~\cite{Soltysiak:2016xqz}.

\subsection{Size of the wave-function}

Let us consider the coupled system $c\bar{c}$-$D^{+}D^{-}$, where $c\bar{c}$
is a charmonium system with quantum numbers $^{3}D_{1}$, and $D^{+}D^{-}$ is
a meson-meson decay channel. The wave function is computed following the
model in Ref.\ \cite{prd94p014016}. For the model parameters
\textquotedblleft string-breaking\textquotedblright\ $4.0$ GeV$^{-1}$ and coupling $0.8$, we
find a pole at $3773.1-i3.4$ MeV, to which corresponds a wave-function with
r.m.s. value $\sqrt{<r^{2}>}=4.74$ GeV$^{-1}$ $\sim 0.93$ fm. If $\sqrt{%
<r^{2}>}\sim 1/\Lambda $, our previously free parameter $\Lambda $ in %
\eqref{cutoff} is around $211$ MeV.

\subsection{Spectral functions\label{sf}}

The self-energy $\Sigma $ of a two-meson loop can be written as 
\begin{equation}
\Sigma (s)=\Omega (s)+i\sqrt{s}\Gamma (s),\ \ \Omega ,\Gamma \in \Re ,
\label{loop}
\end{equation}%
where $\Gamma (s)$ is the width's function  of the resonance and it is given
(see Ref.\ \cite{PDG}) by 
\begin{equation}
\Gamma (s)=\frac{1}{8\pi }\frac{p(s)}{s}|\mathcal{M}|^{2},  \label{gam}
\end{equation}%
while the real part $\Omega $ can be computed from the width through the
Kramers-Kr\"{o}nig dispersion relation 
\begin{equation}
\Omega (s)=\frac{1}{\pi }\int_{s_{th}}^{\infty }\frac{\sqrt{s^{\prime }}%
\Gamma (s^{\prime })}{s^{\prime }-s}\mathrm{d}s^{\prime }.  \label{dr}
\end{equation}%
The propagator is given by 
\begin{equation}
\Delta (s)=\frac{1}{s-m_{\psi }^{2}+\Sigma (s)},  \label{prop}
\end{equation}%
and the spectral function, as a function of the CM energy, by ($E=\sqrt{s}$%
): 
\begin{equation}
d_{\psi }(E)=-\frac{2E}{\pi }\mathrm{Im}\ \Delta (E).  \label{ls}
\end{equation}%
To ensure faster convergence of the integral in Eq.~\eqref{dr} we use,
instead of $\Omega (s)$, the once-subtracted dispersion relation $\Omega
_{1S}(s)=\Omega (s)-\Omega (m_{\psi }^{2})$ leading to $\Sigma
_{1}(s)=\Omega _{1S}(s)+i\sqrt{s}\Gamma (s)$. For further details, see Ref. 
\cite{lupo}.

\subsection{Unitarization}

In the so-called K\"{a}llen-Lehmann representation we have 
\begin{equation}
\Delta (s)=\int_{0}^{\infty }ds^{\prime }\frac{d_{S}(s^{\prime })}{%
s-s^{\prime }+i\varepsilon }
\end{equation}%
in the limit $s\rightarrow \infty $ 
\begin{equation}
\frac{1}{s}=\frac{1}{s}\int_{0}^{\infty }ds^{\prime }d_{S}(s^{\prime
})\Rightarrow \int_{0}^{\infty }ds^{\prime }d_{S}(s^{\prime })=1,
\end{equation}%
where the left part comes from Eq.\ \eqref{prop}, considering that $\Sigma
(s)$ goes to zero, due to the cutoff function.

\subsection{Poles}

In order to find poles on the second Riemann sheet we analytically continue
the loop function Eq.\ \eqref{loop} to the complex plane, and the pole
condition is given when the denominator of the propagator \eqref{prop} is
zero, i.e., 
\begin{equation}
E^2-m_R^2+\Sigma(E)=0,\ E\in\mathbb{C},
\end{equation}
with the energy on the second Riemann sheet.

\section{Line-shape and poles}

\begin{figure}[tbp]
\begin{center}
\resizebox{!}{200pt}{\includegraphics{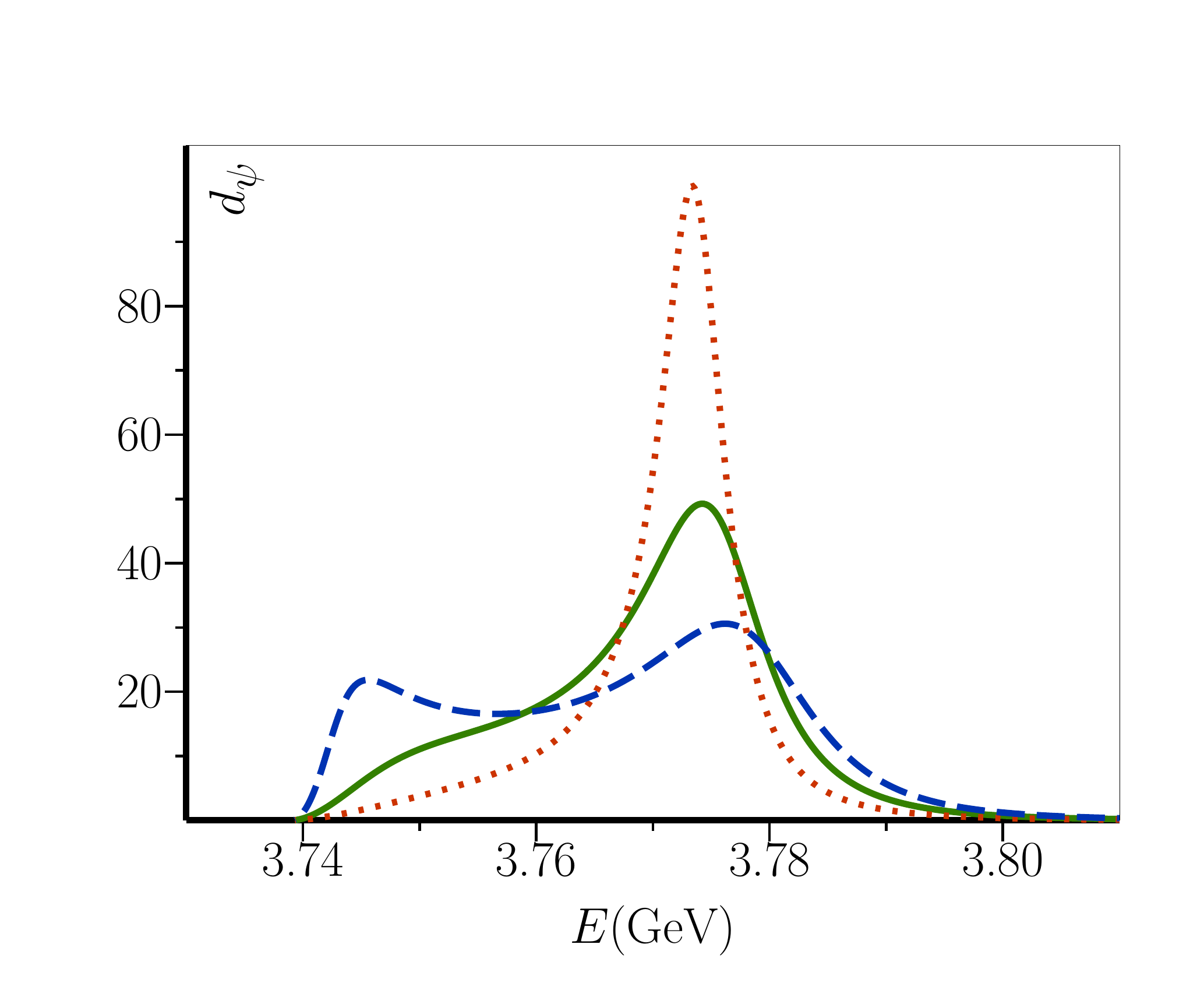}}
\end{center}
\caption{Line-shape of the resonance $\protect\psi (3770)$ in the channel $%
D^{+}D^{-}$. Solid line $\tilde{g}=g$, dotted line $\tilde{g}=0.7g$, and
dashed line $\tilde{g}=1.3g$ (cf.~text).}
\label{gvarls}
\end{figure}

In Fig.~\ref{gvarls} we show the unitarized line-shape distribution,
according to Eq.~\eqref{ls}, in channel $D^{+}D^{-}$, using the parameters $%
m_{\psi }=3773.13$ MeV (mass fit in \cite{PDG}), $\Lambda =211$ MeV, and $%
g_{\psi DD}=44\sqrt{2}$, represented by the solid line. The result
reproduces the structure observed in the BES data, namely the higher slope
on the higher energy side, and the deformation on the lower energy side. We
find two poles corresponding to these parameters: $3744-i11$ MeV and $3775-i6
$ MeV. Furthermore, we study the influence of the strong coupling $g=g_{\psi
DD}$ on the line-shape. For $\tilde{g}=0.7{g}$ the line-shape exhibits only
one peak, yet with two poles at $3741-i20$ MeV and $3778-i3$ MeV. For $%
\tilde{g}=1.3{g}$ the line-shape shows clearly two peaks, corresponding to
the poles $3743-i4$ MeV and $3778-i9$ MeV. The lower energy pole is
generated dynamically and disappears if $g$ is small enough, remaining only
the higher energy pole coming from the \textquotedblleft
seed\textquotedblright . For larger $g$ values, the seed pole moves to
higher energies while the dynamical pole approaches threshold. The existence
of two poles does not necessarily mean the existence of two different
resonances, instead, it means that the pair $D$-$\bar{D}$, more than
contributing to the kinematic background only, plays a dynamical role in the
formation of the $\psi (3770)$. One of the reasons might be not only because
the $\psi (3770)$ is above and close to the $D\bar{D}$ threshold, but also
because it is a dominantly $D-$wave state, and therefore its wave-function
is larger than in case of $S-$wave states, conferring it properties of
lighter systems.

\section{Conclusions and outlook}

A correct understanding of resonance signals is important to disentangle the
nonperturbative phenomena hidden in the line-shapes. Here, we have performed
a dynamical study of $\psi (3770)$ by using an effective Lagrangian model,
which points out the relevance of the $D^{+}D^{-}$ loop, \textit{viz.}%
~coupled-channel, to the formation of the resonance. We find a two-pole
structure in the signal. Further studies include the final state
rescattering, the influence of the cutoff function, and the lepton-lepton
decay widths.

\section*{Acknowledgements}

We thank the organizers for the very nice workshop and M. Piotrowska, T.
Wolkanowski-Gans, G. Rupp, and E. van Beveren for useful discussions. This
work was supported by the \textit{Polish National Science Centre} through
the OPUS project no. 2015/17/B/ST2/01625.

\end{document}